\documentclass[conference]{IEEEtran}

\makeatletter
\def\ps@headings{%
\def\@oddhead{\mbox{}\scriptsize\rightmark \hfil \thepage}%
\def\@evenhead{\scriptsize\thepage \hfil \leftmark\mbox{}}%
\def\@oddfoot{}%
\def\@evenfoot{}}
\makeatother
\usepackage{cite}
\usepackage{amsmath,amssymb,amsfonts}
\usepackage{algorithmic}
\usepackage{graphicx}
\usepackage{textcomp}
\usepackage{xcolor}
\usepackage{balance}
\usepackage{upgreek}
\usepackage{amsmath}
\usepackage{subcaption}
\usepackage{subcaption}
\usepackage{booktabs} 
\usepackage{multirow}
\usepackage{tikz}
\usetikzlibrary{positioning, arrows.meta, fit, calc}

\usepackage{url}
\usepackage{amsmath, amssymb}
\usepackage[most]{tcolorbox}
\usepackage{xcolor}

\usepackage[letterpaper,left=0.625in, right=0.625in, top=0.70in, bottom=1in]{geometry}

\usepackage{xspace}
\newcommand{\IW}{\texttt{Sec5GLoc}\xspace}

\def\BibTeX{{\rm B\kern-.05em{\sc i\kern-.025em b}\kern-.08em
    T\kern-.1667em\lower.7ex\hbox{E}\kern-.125emX}}
\begin{document}

\title{\IW: Securing 5G Indoor Localization via Adversary-Resilient Deep Learning Architecture}

\author{
   \IEEEauthorblockN{
       Ildi Alla\IEEEauthorrefmark{1}, 
        Valeria Loscri\IEEEauthorrefmark{1}
    }
    
    \IEEEauthorblockA{\IEEEauthorrefmark{1}Inria Lille-Nord Europe, France \\
    \{ildi.alla, valeria.loscri\}@inria.fr}
}

\maketitle

\begin{abstract}
Emerging 5G millimeter-wave and sub-6 GHz networks enable high-accuracy indoor localization, but security and privacy vulnerabilities pose serious challenges. In this paper, 
we identify and address threats including \textit{location spoofing} and \textit{adversarial signal manipulation}
against 5G-based indoor localization. We formalize a threat model encompassing attackers who inject forged radio signals or perturb channel measurements to mislead the localization system.
To defend against these threats, we propose an \textit{adversary-resilient localization architecture} that combines deep learning fingerprinting with physical domain knowledge. Our approach integrates multi-anchor Channel Impulse Response (CIR) fingerprints with Time Difference of Arrival (TDoA) features and known anchor positions in a hybrid Convolutional Neural Network (CNN) and multi-head attention network. This design inherently \textit{checks geometric consistency} and \textit{dynamically down-weights anomalous signals}, making localization robust to tampering. 
We formulate the secure localization problem and demonstrate, through extensive experiments on a public 5G indoor dataset, that the proposed system achieves a mean error approximately 0.58\,m under mixed Line-of-Sight (LOS) and Non-Line-of-Sight (NLOS) trajectories in benign conditions and gracefully degrades to around 0.81\,m under attack scenarios. 
We also show via ablation studies that each architecture component (attention mechanism, TDoA, etc.) is critical for both accuracy and resilience, \textit{reducing errors by 4-5 times} compared to baselines. In addition, our system runs in real-time, localizing the user in just 1 ms on a simple CPU. 
The code has been released to ensure reproducibility\footnotemark.

\footnotetext{\url{https://github.com/sec5gloc/Sec5GLoc}}

\end{abstract}

\vspace{-0.05cm}
\section{Introduction}
\vspace{-0.05cm}

Indoor localization has become a cornerstone of modern wireless services, enabling applications from asset tracking in smart factories to navigation in airports and shopping malls. 
The advent of \textit{5G New Radio} (NR) technology, with its large bandwidth (e.g. 100\,MHz or more) and fine time resolution, offers unprecedented opportunities for accurate indoor positioning where Global Positioning System (GPS) is unavailable \cite{chen2021carrier}. High-resolution Channel State Information (CSI) or Channel Impulse Responses (CIR) from 5G signals can serve as detailed “fingerprints” of a user’s location, even under challenging Non-Line-of-Sight (NLOS) conditions. Recent research has shown that deep learning models, such as Convolutional Neural Networks (CNNs), can leverage these rich 5G signal features to achieve sub-meter localization accuracy, outperforming classical methods in complex indoor environments \cite{zhang2023csi}.

However, the \textit{security and privacy} of such 5G-based indoor localization systems have received far less attention. Unlike outdoor Global Navigation Satellite Systems (GNSS), which has known spoofing issues, indoor localization inherits unique vulnerabilities from wireless networks. The \textit{open nature of radio channels} means that an adversary can inject or manipulate signals in the physical layer – attacks that bypass conventional cryptographic protections \cite{10917719}. For example, an attacker could place absorptive or reflective materials around devices to distort received signal strengths \cite{li2011designing}, or even impersonate legitimate 5G base stations (anchors) using software-defined radios (SDRs) \cite{mubasshir2024fbsdetector}. Such \textit{non-cryptographic attacks} on the signal measurements can lead to large localization errors or completely false position estimates while remaining invisible to authentication protocols. These threats are not hypothetical: prior work in Wi-Fi and sensor networks demonstrated that carefully \textit{spoofed signals} can fool fingerprint-based positioning systems, yielding attacker-chosen locations or denial-of-service. For instance, the authors in \cite{yuan2018secure} showed that an attacker who learns the radio-map of a Wi-Fi positioning system can then transmit \textit{fake access point signals} with tuned power to mislead the location server by meters. Similarly, the authors in \cite{tippenhauer2009attacks} exploited the lack of physical-layer security in WLAN to launch \textit{public Wi-Fi localization attacks}, and in \cite{bauer2009directional} demonstrated a “\textit{directional antenna attack}” using a tin can to spoof location estimates. These studies underscore that indoor localization systems are inherently vulnerable to \textit{signal spoofing attacks} where adversaries mimic legitimate patterns to deceive the system \cite{han2025rf}.

In addition to spoofing and tampering attacks that threaten localization \textit{integrity}, 5G positioning raises new \textit{privacy and inference} concerns. In emerging 5G deployments, positioning is often \textit{network-centric} (“off-device”), meaning the infrastructure (5G base stations or location servers) computes the user’s location using uplink channel measurements. This approach can \textit{unintentionally broadcast a user’s location information} to any sufficiently equipped eavesdropper. As the authors in \cite{huang2024attacking} observe, any device in range that can capture the user’s 5G CSI can potentially infer the user’s position, creating serious privacy risks. Conversely, a malicious user can exploit the system to infer sensitive environment data – e.g., repeatedly querying a localization service to \textit{learn the fingerprint database or model}, effectively performing a model-inversion attack \cite{yuan2018secure}. Such \textit{inference attacks} could reveal floor plans or device locations that should remain confidential. Moreover, the incorporation of machine learning (ML) introduces the possibility of \textit{adversarial attacks}. An attacker might craft subtle perturbations to the input signals (e.g., adding carefully timed multipath reflections) that cause the deep learning model to output arbitrary locations, all while standard performance appears normal. This is analogous to adversarial examples in computer vision, but in the radio frequency (RF) domain where generating them requires controlling the wireless propagation environment \cite{han2025rf}. 

\textit{Securing 5G indoor localization} is therefore an urgent challenge. A secure localization system must provide \textit{accurate positioning despite adversarial interference}, while preventing unauthorized disclosure of location information. Traditional approaches to secure localization, developed for earlier wireless systems, only partially address these needs. For example, authors have proposed robust statistical methods and anomaly detectors to identify outlier measurements in localization \cite{li2005robust, weber2020gordian}, or cryptographic distance-bounding protocols to thwart impersonation in ultra-wideband systems (UWB) \cite{luo2024secure}. These defenses often assume a limited attacker (e.g. only a minority of signals under attack \cite{li2011designing}) and may not scale to the rich, high-dimensional input of a 5G CSI fingerprint. Recent studies have begun exploring machine learning-based defenses. For instance, the authors in \cite{gufran2024sentinel} introduce capsule networks to increase resilience against \textit{rogue anchor attacks}, achieving three times lower error under attack compared to prior models. Likewise, Generative Adversarial Networks (GANs) have been used to \textit{detect and remove adversarial perturbations} from Wi-Fi fingerprints \cite{yan2024transgan}. These works point toward the need for \textit{hybrid solutions} that combine the strengths of data-driven learning and physics-based safeguards.

In this paper, we present a \textit{security-enhanced 5G indoor localization architecture} that addresses the above challenges. 
By considering the state-of-the-art deep learning-based fingerprinting, we develop a new approach that integrates  \textit{security and privacy} features by design.
Our approach incorporates a formal threat model and introduces architectural features to defend against \textit{spoofing and adversarial manipulation}. Specifically, we leverage the known geometry of anchors and timing information as \textit{built-in consistency checks}, and we use a \textit{multi-head attention} fusion mechanism to dynamically mitigate the impact of suspicious signals. We also consider the system’s deployment model to enhance privacy (e.g. enabling on-device localization to keep CSI data local).

\smallskip
\noindent \textbf{Summary of Novel Contributions}\smallskip

\noindent $\bullet$ We propose a novel deep learning-based localization architecture that is intrinsically resilient to many signal attacks. The architecture, called \IW (Secure 5G Localization), integrates a CNN-based fingerprinting model with \textit{physical features} (anchor positions and TDoA) and an \textit{attention-based multi-anchor fusion} mechanism. By combining learned RF fingerprinting with model-driven checks, \IW achieves robust performance even when some \textit{inputs are compromised}.

\noindent $\bullet$ We formulate the secure localization problem as a robust \textit{regression task} under adversarial perturbations. We show how incorporating known anchor coordinates and inter-signal time differences acts as a defense by constraining the model to respect the speed-of-light geometry, thereby raising the bar for attackers. We discuss how the attention mechanism can serve as an implicit \textit{anomaly detector} by reducing the weight of out-of-pattern anchor measurements.

\noindent $\bullet$ We evaluate our approach on a public 5G indoor localization dataset \cite{stahlketrack}, collected in a large industrial environment under different conditions. \IW achieves mean errors of 0.34\,m in NLOS areas and 0.58\,m in mixed LOS and NLOS areas, despite not being trained on LOS areas. Compared to classical $k$-NN fingerprinting and geometric TDoA multilateration, our approach \textit{reduces error by over 75\% in mixed conditions}. We further simulate robustness against spoofing attack, noise perturbation, and anchor drop scenario—each applied to the \textit{most influential anchor}—showing only slight performance degradation, thanks to the attention-based re-weighting of anchors. In addition, we perform ablation studies showing that removing attention or physical features \textit{increases error by 2–3 times} and breaks generalization, confirming each component's role in accuracy and resilience.


\vspace{-0.05cm}
\section{Related Work}
\vspace{-0.05cm}
\subsection{Secure Indoor Localization} 
\vspace{-0.05cm}
Security issues in RF localization have been studied for over two decades. Early work identified that traditional localization algorithms (e.g., those based on signal strength or time-of-flight) are vulnerable to \textit{physical-layer attacks} that cannot be prevented by cryptography \cite{yuan2018secure}. The authors in \cite{li2011designing} conducted one of the first comprehensive analyses of localization under signal strength attacks, showing that an adversary can attenuate or amplify signals to significantly bias position estimates. 
Follow-up research proposed various countermeasures.
For instance, \textit{robust statistics} were introduced to filter out extreme Received Signal Strength (RSS) readings caused by attackers \cite{li2005robust}. The authors in \cite{li2011designing} developed a \textit{ratio consistency check} that uses differences of signal strength rather than absolute values, to cancel out uniform attenuation introduced by attackers. In wireless sensor networks, secure localization algorithms like \textit{verifiable multilateration} and \textit{distance bounding} were developed, where nodes exchange challenge-response signals to ensure the measured distance is authentic \cite{capkun2005secure}. The work
showed that by combining distance bounding with multiple reference points, devices can compute a position that is robust against attackers who delay or replay signals.  However, these methods often require specialized hardware or tight time synchronization.

Another line of work focuses on \textit{detecting and mitigating spoofing} in fingerprinting systems. Traditional RSS fingerprint systems, such as the classic RADAR system \cite{yang2012detection}, assume a static radio map. If an attacker introduces a \textit{rogue access point} or alters the environment, the fingerprint matching process may produce errors. Methods like \textit{K-means clustering of fingerprints} have been used to detect inconsistent readings. The authors in \cite{yuan2018secure} specifically examined active signal spoofing on RSS-based indoor positioning. They identified two practical attacks: one where an attacker \textit{learns the fingerprint database by querying}, and another where they \textit{impersonate Wi-Fi APs with controlled power} to manipulate a target’s observed RSS. Their proposed countermeasure used a truncated distance metric to make fingerprint matching less sensitive to the attacker’s changes, which reduced but did not fully eliminate the attack impact. These efforts indicate that securing fingerprint-based localization is challenging – defenses tend to be scenario-specific and can be overcome by more sophisticated adversaries (e.g., those who control many signals or craft perturbations that evade simple filters).

\vspace{-0.05cm}
\subsection{Adversarial Attacks on Wireless ML Systems}
\vspace{-0.05cm}

With the rise of machine learning for wireless sensing and localization, researchers have begun examining how \textit{adversarial machine learning} techniques transfer to the wireless domain. Traditional adversarial examples involve adding small perturbations to sensor inputs (images, etc.) to fool a neural network. In wireless localization, the “input” is the set of RF signals (channel measurements) observed. Several works have demonstrated feasibility of such attacks.

The work in \cite{huang2024attacking} consider a \textit{deep learning-based positioning system} where the network infrastructure computes a device’s location from its uplink CSI. They show that a malicious device can transmit carefully crafted signals – for example, by altering the phase or amplitude of certain Orthogonal Frequency-Division Multiplexing (OFDM) subcarriers – to \textit{mislead the position estimator while remaining protocol-compliant}. This is essentially an \textit{adversarial example over-the-air}. By evaluating on real 5G and Wi-Fi CSI datasets, they demonstrated that an attacker can cause significant localization errors with minimal impact on communication quality, thereby staying stealthy. Their work also explored defenses like adversarial training, finding that improving robustness often comes at a cost of localization accuracy or system complexity.

The authors in \cite{gufran2024sentinel} address adversarial attacks on \textit{Wi-Fi RSS fingerprinting} with their system \textit{SENTINEL}. They highlight the threat of \textit{rogue anchors (APs)} broadcasting misleading signals to confuse ML models. SENTINEL employs modified capsule networks, which are reputed to be more robust to perturbations, and introduces a new dataset (RSSRogueLoc) with real-world rogue AP scenarios. Experiments show a \textit{3.5 times reduction in mean error under simulated attacks} compared to prior deep models, illustrating that ML architectures can be made more attack-resilient. Another study by \cite{yan2024transgan} uses a GAN to preprocess RSS fingerprints and \textit{remove adversarial noise}. By training a generator to produce “clean” fingerprints from attacked ones, they improved the robustness of a fingerprinting system without significantly hurting its baseline accuracy.

A related threat is using the localization system itself as an oracle to infer information.
For example, \textit{WindTalke} \cite{li2016csi} showed that by sniffing CSI from Wi-Fi, an attacker could infer keystrokes typed by a victim, representing a side-channel attack on localization-related signals. While not directly about position, it exemplifies how subtle information in RF signals can be abused by adversaries. In the context of location systems, an attacker may exploit changes in reported locations or confidence scores to deduce if a user is present in a certain area (tracking), or as mentioned earlier, may query a service to reconstruct its radio map. Defenses here include techniques like \textit{privacy-preserving localization} – e.g., adding noise to location outputs, secure multi-party computation for localization, or limiting the granularity of public location queries. An example is the work by \cite{li2018secure} on \textit{secure crowdsourced indoor positioning}, which considered dishonest participants and proposed safeguards when building a location service from user-contributed data.

\vspace{-0.05cm}
\subsection{5G Positioning and Security}
\vspace{-0.05cm}

Specific to 5G systems, the standards themselves incorporate some location security features, but also new attack surfaces. The 5G Positioning Protocol (LPP) and the Location Management Function (LMF) can use \textit{encryption and authentication} to protect location reports. Yet, at the physical layer, Positioning Reference Signals (PRS) and sounding signals are typically unauthenticated. An attacker could potentially \textit{broadcast fake PRS} if they have comparable hardware.
A recent overview by \cite{abuyaghi2024positioning} enumerates possible attack targets in the 5G positioning ecosystem, including the LMF, the interface between gNodeBs and the location server, and the PRS transmissions themselves. Threats range from \textit{jamming PRS (denial-of-service) to man-in-the-middle attacks} that alter location results delivered to the user or application. While our focus is on the RF/ML side, these system-level threats are important to note. \textbf{A fully secure solution would need to secure both the physical measurements (our focus) and the network infrastructure (beyond scope here).}


\vspace{-0.05cm}
\section{Threat Model}
\vspace{-0.05cm}

We consider an indoor localization system where a \textit{set of trusted 5G anchors} (base stations or access points) with known positions provide signals used for positioning a mobile device. The device’s goal is to determine its own location or have the network to determine it based on the received radio signals. Without loss of generality, we assume an \textit{uplink TDoA scenario} – the device transmits a sounding signal that is received by multiple anchors, each measuring a CIR or timing, and a location server fuses this information to compute the device’s coordinates. Our approach also applies to a downlink fingerprinting scenario, since we ultimately use the CIRs from multiple anchor-device links as input features.

\noindent \textbf{Adversary Capabilities.}~We assume the adversary \textit{does not compromise the cryptographic or higher-layer security} of the 5G network. For instance, they do not steal encryption keys or hijack the location server. Instead, the adversary operates at the physical layer and the usage interface. The following capabilities are considered:

\noindent $\bullet$ \textit{Signal Injection (Spoofing).}~The attacker can inject radio signals that mimic those used for localization \cite{han2025rf}. For example, the attacker may deploy a rogue transmitter in the environment that emits a 5G-like reference signal or echo. This could be done by spoofing an anchor, pretending to be an additional base station, or by replaying the device’s signal with delays. The attacker has significant control over the timing, power, and perhaps phase of these signals, allowing them to create \textit{false distance measurements}. However, we assume the attacker is constrained by \textit{physical realism}. They cannot, for instance, instantaneously create a fake signal that arrives earlier (faster than light) than the true direct path. They also may be constrained in power, as a high-power attack would be easily detectable or would disrupt communication. Therefore, we consider stealthier attacks where the perturbations are 
subtle and remain within a similar magnitude to normal signals.

\noindent $\bullet$ \textit{Environmental Manipulation.}~The attacker can place objects or reflectors to modify propagation. This includes moving metallic objects, blocking certain paths, or introducing reflectors that create additional multipath. Such manipulation can \textit{bias the CIR fingerprints or TDoA}. For instance, placing a metal sheet could remove a path or delay it, imitating an increased distance. Unlike direct signal injection, this indirect method might be harder to attribute to an attack, as it could appear as natural environmental change. We consider this as part of adversarial signal manipulation.



\noindent $\bullet$ \textit{Denial-of-Service (DoS).}~Although our focus is on deception 
attacks, we acknowledge that an attacker could simply jam the signals, preventing the system from functioning. Jamming is a trivial but effective attack on any wireless system. We don't consider jamming as the primary threat in this work, as it is a pure availability attack and can often be mitigated by spread-spectrum or detection of lost signals. Our emphasis is on \textit{subtle attacks that degrade integrity without obvious detection}.



\noindent \textbf{Adversary Goals.}~The attacker wants to cause the system to output an incorrect location for the target device. This could mean a \textit{large error}, making the localization unusable or dangerous, or a \textit{specific spoofed location} (e.g., making a device appear in a different room). For example, an attacker might try to make a security robot believe it is at a different location, or cause a tracking system to lose an asset. In technical terms, the attacker aims to maximize the localization error or achieve a particular offset, while remaining undetected by any anomaly checks.


\noindent \textbf{Defender Assumptions.}~Our system assumes the anchors and the location server are \textit{trusted and secure}. Anchors are time-synchronized and share their measurements honestly. We assume the attacker has not compromised these infrastructure elements (i.e., no insider at anchor). Communication from anchors to the server can be secured by traditional means (encryption, authentication), so we focus on attacks that \textit{happen before or during} signal measurement. We also assume the attacker cannot infinitely overpower the signals without being noticed. Extremely strong interference would likely trigger alarms or at least degrade communication, which is outside the attacker’s interest if stealthy manipulation is the goal. 
\textbf{The attacker's perturbations are thus bounded but not necessarily small, potentially significant yet not blatantly obvious.}


\vspace{-0.05cm}
\section{Problem Formulation}
\begin{figure}[ht]
    \centering
\includegraphics[width=0.285\textwidth]{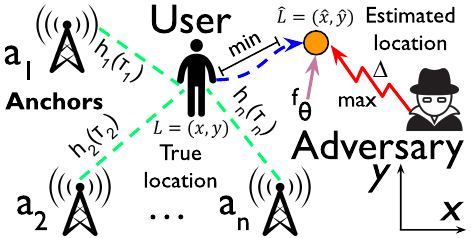}
    \vspace{-0.2cm}
    \caption{Problem formulation of secure and robust 5G localization under adversarial signal perturbations and physical-layer threats.
}
    \label{fig:problem formulation}
    \vspace{-0.31cm}
\end{figure}

\vspace{-0.05cm}
We formulate secure indoor localization as a \textit{robust regression problem} under adversarial conditions (Figure~\ref{fig:problem formulation}). Let $N$ be the number of anchors (reference points) deployed, each at a known 2D/3D coordinate $\mathbf{a}_i$ for $i=1,\dots,N$. At a given time, a device at true location $\mathbf{L}=(x,y)$ transmits a signal. Anchor $i$ receives a waveform which can be characterized by its CIR $h_i(\tau)$ as a function of delay $\tau$. In an ideal noise-free scenario without attackers, $h_i(\tau)$ has a distinct peak at the propagation delay $\tau_i = |\mathbf{L}-\mathbf{a}_i|/c$ (distance divided by signal speed), and possibly additional multipath components at longer delays due to reflections. The TDoA between any two anchors $i$ and $j$ in line-of-sight is $\tau_i - \tau_j$, which geometrically constrains $\mathbf{L}$ to a hyperbola. In practice, due to multipath in NLOS, the direct peak may be obscured, and machine learning is used to infer $\mathbf{L}$ from the \textit{pattern} of $h_i(\tau)$ across all anchors.

\vspace{-0.31cm}

We denote by $\mathbf{H} = \{h_i(\tau)\}_{i=1}^N$ the collection of signals from all anchors. \textit{Our baseline localization function (without security) is a deep network} $f_\theta$ (with parameters $\theta$) that takes $\mathbf{H}$ or a suitable feature representation of it and outputs an estimate $\hat{\mathbf{L}} = f_\theta(\mathbf{H}, \{\mathbf{a}_i\}_{i=1}^N)$. Here, \textit{we explicitly allow} $f_{\theta}$ to depend on the known anchor positions ${\mathbf{a}_i}$ as inputs, which will be important for generalization. The model is trained on examples of $(\mathbf{H}, \mathbf{L})$ from the environment.

\noindent \textbf{Adversarial Modeling.}~The adversary can modify the received signals to $\tilde{\mathbf{H}} = { \tilde{h}_i(\tau)}$ before they are processed by the localization algorithm. We capture the attacker’s capabilities in a \textit{perturbation function} $\Delta$ applied to the true signals: $\tilde{\mathbf{H}} = \Delta(\mathbf{H}, \mathcal{A})$, where $\mathcal{A}$ represents the \textit{attacker’s actions/strategy} (e.g., inserted signals, delays, etc.). For example, $\Delta$ might add an extra path of a certain amplitude at a chosen delay to some $\tilde{h}_i$. We assume $\Delta$ is bounded in a sense of \textit{not creating physically impossible patterns}. One bound is that $\tilde{h}_i(\tau)$ must have no significant energy at negative delays or much earlier than the line-of-sight would allow. 
Another bound could be that the total power of injected signals is less than or equal to $\epsilon$ times the legitimate signal power (an “attack budget”). However, we keep the formulation general.

\noindent \textbf{Objective Formulation.}~The secure localization problem can then be stated as: \textit{find an estimator $f$ that minimizes the localization error under the worst-case perturbation up to the allowed budget}. Formally, we want to design $f$ to minimize: 

\vspace{-0.2cm}
\begin{tcolorbox}[
    enhanced,
    coltitle=white,
    colback=gray!10,
    colframe=black,
    fonttitle=\bfseries,
    sharp corners,
    boxrule=0.8pt,
    top=1pt, bottom=1pt, left=2pt, right=2pt,
    boxsep=1pt,  
    title={},    
    before upper={\vspace{1.1em}},
    overlay={
        \node[anchor=north west, inner sep=2pt, outer sep=0pt, font=\bfseries, text=white, fill=gray!98, draw=black]
        at ([xshift=0.8pt,yshift=-0.8pt]frame.north west)
        {General Problem Definition};
    }
]
\begin{equation}
\begin{aligned}
    &\min_{f} \max_{\Delta} \quad \left\| f(\Delta(\mathbf{H})) - \mathbf{L} \right\|, \\
    &\text{subject to} \quad \Delta \in \mathcal{D},
\end{aligned}
\tag{1}
\end{equation}
\end{tcolorbox}
\vspace{-0.2cm}

\noindent where $\mathcal{D}$ is the set of admissible attack perturbations determined by the threat model. This is akin to a \textit{min-max optimization}, which in practice we address through design rather than solving directly.

We also consider detection: ideally, if an attack is too large to correct, the system should detect an anomaly. While our primary objective is robust estimation, we will note how certain intermediate outputs, such as attention weights, could signal an attack. Another aspect is privacy: if the system is off-device, an eavesdropper could estimate $\mathbf{L}$ from $\mathbf{H}$ as well. Privacy preservation would mean \textit{limiting the information leaked by $\mathbf{H}$ to unauthorized parties}. Solutions might involve encrypting the reference signals or obfuscating the CSI. In this paper, we focus on robustness. We assume privacy can be handled by deploying our model on the device or in a secure enclave so that $\mathbf{H}$ is not broadly broadcast.

In summary, our goal is to create a function $f(\cdot)$ that produces accurate $\hat{\mathbf{L}}$ for all $\tilde{\mathbf{H}}$ in a realistic attack model. We incorporate domain knowledge (anchor positions, time-of-flight physics) into $f$ to restrict the space of mistakes it can make. The next section describes the architecture of $f$, i.e., our proposed \IW deep learning model, and how it is trained to resist attacks.

\vspace{-0.05cm}
\section{Proposed \IW Architecture}
\label{Proposed architecture}
\vspace{-0.05cm}

\begin{figure}[ht]
    \centering
\includegraphics[width=0.4\textwidth]{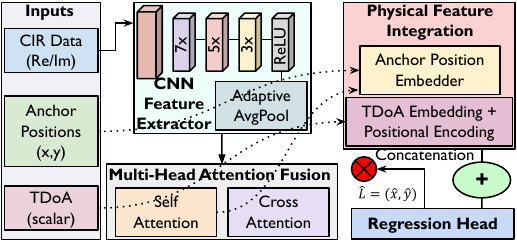}
    \vspace{-0.15cm}
\caption{\IW model: CIR features are extracted using shared CNN; anchor positions and TDoA features are embedded and fused via attention, then passed to a regression head to predict the location.}

    \label{fig:proposed architecture}
    \vspace{-0.2cm}
\end{figure}

We design a \textit{hybrid architecture} that blends deep learning with physical modeling to achieve both high accuracy and resilience to attacks. 
The core of our system is a deep neural network that we call \IW, illustrated at a high level in Figure~\ref{fig:proposed architecture}.
The architecture consists of four main modules: (a) \textit{the CNN feature extractor for CIRs}, (b) \textit{the multi-head attention fusion}, (c) \textit{the physical feature (anchor positions and TDoA) integration}, and (d) \textit{the  regression head}. 
We describe each component and highlight how it contributes to security.

\vspace{-0.2cm}

\noindent \textbf{(a) CNN Feature Extractor per Anchor.}~Each anchor’s CIR $h_i(\tau)$ is first converted into a suitable input tensor for the neural network. We represent the complex CIR as two separate channels: \textit{real and imaginary components}. This forms a tensor of size (2 × $L$), where $L$ is the length of the CIR in time samples. A CNN is applied to each anchor’s data independently to extract high-level features. The CNN is shared (\textit{same weights}) across all anchors, which enforces that features are extracted in a consistent manner. The CNN have several layers of 1D convolutions 
($\tau$ axis is one-dimensional) and pooling, capturing patterns like the presence of direct path, energy in certain delay bins, etc. By training on many examples, the CNN learns to emphasize features that correlate with position (e.g., the “\textit{fingerprint}” of a multipath unique to a location).

\noindent \textbf{\textit{Security advantage:}}~The CNN itself is primarily for accuracy, but a subtle benefit is that it compresses the raw data into a feature vector. This could potentially discard high-frequency noise, making the system a bit robust to minor perturbations. More importantly, sharing CNN weights across anchors means \textit{any anomalies in one anchor’s signal will show up as an outlier} in feature space, which the next stage can catch.

\noindent \textbf{(b) Multi-Head Attention Fusion.}~We obtain a feature vector $f_i$ from each anchor $i$’s CNN. Now, we need to \textit{fuse information} from all anchors to estimate the location. A naive approach is concatenation or averaging, but we introduce a \textit{multi-head attention mechanism} to perform a weighted fusion. In essence, the attention module learns to assign a weight (attention score) to each anchor’s features for the task of localizing the current sample. The module is multi-head, meaning it can compute \textit{multiple sets of weights} focusing on different aspects of the features, which are then combined.

Concretely, we can think of the anchor features ${f_i}$ as a sequence. The attention computes something like:
\vspace{-0.2cm}
\begin{equation}
    \alpha_i = \text{Attention\ Weight}\left(f_i, \left\{ f_j \right\}_{j=1}^N \right),
    \vspace{-0.2cm}
\end{equation}
with $\sum_i \alpha_i = 1$ after softmax. Then, the fused representation is $F = \sum_i \alpha_i \cdot W_f f_i$, using linear projection $W_f$. Multi-head means we do this $M$ times with different learned parameters and concatenate the results.

\noindent \textbf{\textit{Security advantage:}}~The attention mechanism \textit{adapts to input anomalies}. In normal conditions, attention will learn to emphasize the anchors that provide the most information for localization (e.g., perhaps an anchor that currently has a strong LOS signal).  If one anchor’s data is inconsistent (say an attacker injected a false multipath causing its feature $f_k$ to deviate), the attention network can learn to give that anchor a lower weight. In effect, the system can \textit{down-weight corrupted anchors}, similar to how robust statistics would reject outliers, but learned automatically. This is particularly powerful when not all anchors are attacked – as long as a majority provide honest information, attention can focus on them. Our ablation results will show that without attention, a single bad anchor can drag the result more significantly. The multi-head aspect \textit{allows capturing different attack patterns}. For instance, one head might focus on overall signal strength to catch outright power anomalies, while another focuses on timing consistency.

Additionally, attention provides \textit{interpretability}: by monitoring the attention scores ${\alpha_i}$ at runtime, the system operators could detect if one anchor is consistently being ignored by the model – a potential sign that something is wrong with that anchor’s signal, maybe an attack or hardware fault. Thus, it doubles as an anomaly indicator.

\noindent \textbf{(c) Integration of Anchor Positions and TDoA features.}~A unique aspect of our design is feeding the known anchor geometry into the model’s middle layers. We  concatenate each anchor’s coordinate $\mathbf{a}_i = (x_i, y_i)$ with its CNN feature $f_i$. We also explicitly compute a time feature: e.g., the \textit{relative arrival time} of that anchor’s peak signal compared to a reference anchor’s peak. These TDoA-inspired features give a coarse sense of distance differences. The coordinates and time features are processed, possibly through a small neural layer, alongside the attention module. Intuitively, this gives the model some \textit{physical context}: it knows where anchors are, so it can learn physics rules like “if anchor A (north side) got the signal much later than anchor B (south side), the user is likely closer to B.” Such hints greatly aid generalization.

\noindent \textbf{\textit{Security advantage:}}~By incorporating physics (positions and TDoA), we \textit{constrain the model with geometry}. This means the model’s output is influenced not just by pattern matching of fingerprints, but by \textit{an understanding of propagation}. From a security perspective, this makes certain spoofing attacks harder. For example, if an attacker injects a fake early signal to anchor A to imply the device is near A, the model, knowing anchor positions, might flag that “anchor A is on the opposite side of the others, it’s unlikely the device moved so much that others still see consistent patterns.” In other words, \textit{the model is less of a black box} – it won’t produce a location that defies the anchor layout and timing. Our results indeed showed that without position/TDoA inputs, the model suffered on scenarios outside training, like open LOS areas, indicating over-reliance on fingerprint memorization. With these inputs, it became more \textit{robust to environmental change}, which analogously helps against malicious changes.

\noindent \textbf{(d) Output Regression and Training.}~The final fused representation, derived from attention and additional fully-connected layers, is fed into a regression output that predicts $(\hat{x}, \hat{y})$. We train the network on a large dataset of collected CIRs with ground-truth positions, using a loss such as \textit{Huber} \cite{barron2019general}, which balances the mean squared error (MSE) and mean absolute error (MAE), offering robustness to outliers while maintaining differentiability around zero. Specifically, if $\mathbf{p} = (x, y)$ is the ground truth position and $\hat{\mathbf{p}} = (\hat{x}, \hat{y})$ is the predicted position, the loss is computed as:
\vspace{-0.25cm}
\begin{equation}
\mathcal{L}(\hat{\mathbf{p}}, \mathbf{p}) = 
\sum_{j \in \{x, y\}}
\begin{cases}
0.5 \cdot (\hat{p}_j - p_j)^2, & \text{if } |\hat{p}_j - p_j| < 1 \\
|\hat{p}_j - p_j| - 0.5, & \text{otherwise}
\end{cases},
\vspace{-0.25cm}
\end{equation}
where the absolute value and squared operations are applied element-wise to the $(x, y)$ coordinates. We also include during training a variety of conditions (different orientations, environment changes) so that the model doesn’t overfit to one fingerprint scenario.

To incorporate adversarial robustness in training, one could employ \textit{adversarial training} by generating simulated attacks on the fly and training the model to handle them. In this work, we did not fully integrate adversarial training due to the complexity of modeling physical attacks. However, we did augment the training with some “\textit{difficult}” cases (e.g., NLOS signals where direct path is absent) to force the model to rely on other anchors.  In the future, one could extend the training with known attack patterns, such as adding synthetic multipath, to further harden the model.

\noindent \textbf{Summary of Defense Mechanisms.}~Our secure architecture provides defense in depth: the \textit{multi-anchor design} means no single measurement determines the outcome; the \textit{attention mechanism} dynamically identifies and down-weights suspicious measurements; and the \textit{geometry-aware features} ensure outputs remain physically plausible, reducing attack degrees of freedom. Importantly, these defenses are achieved without additional external modules – they are part of the learning model, which keeps the solution efficient. The model contains on the order of $10^5$ parameters and can \textit{run in real-time}. 
This real-time capability means it could potentially check each incoming localization in an online system, making it practical for continuous monitoring or tracking applications.


\begin{table*}[t]
\centering
\caption{Performance of \IW versus baselines in benign conditions.}
\vspace{-0.2cm}
\resizebox{\linewidth}{!}{
\begin{tabular}{llcccc}
\toprule
\textbf{} & \textbf{Method (Scenario)} & \textbf{$Mean \ Error$} & \textbf{$Median \ Error$} & \textbf{$75^{th}\ Percentile \ Error$} & \textbf{$90^{th} \ Percentile \ Error$} \\
\textbf{} & \textbf{} & ($m$) & ($m$) & ($m$) & ($m$) \\
\midrule
\multirow{1}{*}{Main Results} 
    & \IW (NLOS) & 0.34 & 0.28 & 0.43 & 0.61 \\
    & \IW (LOS and NLOS)  & 0.58 & 0.46 & 0.75 & 1.19 \\
\midrule
\multirow{3}{*}{Baselines} 
    & $K$-NN Fingerprinting (NLOS)      & 0.67 & 0.13 & 0.75 & 1.99 \\
    & $K$-NN Fingerprinting (LOS and NLOS)      & 3.31 & 2.67 & 4.49 & 7.45 \\
        & Classical TDoA (NLOS)   & 2.39 & 1.97 & 3.11 & 4.59 \\
    & Classical TDoA (LOS and NLOS)          & 2.25 & 1.83 & 2.94 & 4.31 \\

\bottomrule
\end{tabular}}
\vspace{-0.5cm}
\label{tab:loc-results}
\end{table*}

\vspace{-0.05cm}
\section{Experimental Evaluation}
\vspace{-0.05cm}
We evaluate the proposed \IW system on a challenging indoor 5G localization dataset and compare it against baseline methods. We focus on two aspects: \textbf{(i)} baseline accuracy in diverse conditions to ensure our security additions did not degrade performance, and \textbf{(ii)} robustness in scenarios that mimic attacks or new environments to validate security benefits.

\noindent \textbf{Dataset and Environment.}~We use the publicly available dataset from \cite{stahlketrack}, which provides uplink 5G NR CIR measurements collected in a large industrial warehouse spanning approximately 1,200\,m\textsuperscript{2}. The environment consists of interconnected halls, reflective metal walls, shelving units, and partially open spaces, with industrial vehicles and metal structures introducing complex multipath effects. There are $N=8$ anchors (5G base station equivalents) strategically installed at known coordinates and varying heights along the perimeter of the environment. The dataset includes \textit{training trajectories} where a device was carried through predominantly NLOS routes, with signals frequently obstructed by shelving and metal objects, and \textit{testing trajectories} that encompass both NLOS and previously unseen open LOS areas to evaluate generalization capabilities. Ground truth positions with sub-decimeter accuracy are provided via a tracking system for evaluation.

\noindent \textbf{Baselines.}~We compare against two main baselines:

\noindent \tikz[baseline=(char.base)]\node[shape=circle, fill=black, inner sep=0.8pt, text=white] (char) {1}; \textit{$K$-NN Fingerprinting.}~A classical fingerprinting method: store all training CIRs with their corresponding positions as reference fingerprints in a database. To localize a new CIR, we extract statistical and signal-based features for each anchor. These features are concatenated across all anchors to form a fixed-dimensional representation of the CIR. During training and validation, we compute distances (Euclidean, Manhattan or Minkowski) in this feature space to all training fingerprints. Localization is performed by finding the $k$-nearest neighbors (we experimented with $k=3, 5, 7,  \dots$) and averaging their coordinates to estimate the position. A grid search over $k$, distance metrics, and weighting schemes (uniform versus distance-weighted) is conducted during validation to identify the best-performing configuration. 
This baseline represents a non-learning-based approach that leverages spatial similarity for localization.

\noindent \tikz[baseline=(char.base)]\node[shape=circle, fill=black, inner sep=0.8pt, text=white] (char) {2}; \textit{Geometric TDoA Multilateration.}~We implement a least-squares multilateration approach. From each burst, we extract the TOA of the first path per anchor. Using the anchor with the minimum TOA as the reference, we compute $\Delta t_i = TOA_i - TOA_{\text{ref}}$ for each anchor $i$. Then, we solve the system of nonlinear equations:
\vspace{-0.12cm}
\begin{equation}
    |\mathbf{p} - \mathbf{a}_i| - |\mathbf{p} - \mathbf{a}_{\text{ref}}| = c \cdot \Delta t_i,
    \vspace{-0.2cm}
\end{equation}
where $\mathbf{p}$ is the unknown transmitter location, $\mathbf{a}_i=(x_i,y_i)$ are anchor coordinates, and $c$ is the speed of light. We solve this using a nonlinear least-squares optimizer. In LOS settings, this method can be accurate, but in NLOS conditions, TOA biases (due to excess delay) lead to significant errors. We include this baseline to quantify the degradation caused by multipath and NLOS propagation. 

\noindent \textbf{Localization Metrics.}~We report standard localization error metrics: mean error, median error, $75^{th}$ percentile error ($p75$), and $90^{th}$ percentile error ($p90$). 
For robust performance, we pay attention especially to $p90$. A lower $p90$ means that even in the worst 10\% cases, the error isn’t too large, which is relevant for security because an attacker might try to cause worst-case errors.

\vspace{-0.05cm}
\subsection{Localization Accuracy in Benign Conditions}
\vspace{-0.05cm}
First, we verify that \IW achieves state-of-the-art accuracy on the given dataset without any attacker interference. Table~\ref{tab:loc-results} summarizes the error statistics. Our model achieves a mean error of 0.58\,m and $p75$ error of 0.75\,m on the test set that includes a mix of NLOS and LOS areas. In pure NLOS sections, performance is even better ($p75$ error: 0.43\,m). This represents a significant \textit{improvement} over both baselines: \textbf{\textit{i)}} the $k$-NN fingerprinting yields a mean error of about 0.67\,m in NLOS conditions but struggles particularly in unseen LOS regions, resulting in mean error of 3.31\,m; \textbf{\textit{ii)}} the classical TDoA approach achieves less than 1\,m error in true LOS but completely \textit{fails} in heavy NLOS areas ($p90$ error: 4.59\,m). In other words, our hybrid approach successfully bridged the gap, performing well across mixed conditions where either baseline alone collapses.



\vspace{-0.25cm}
\begin{figure}[ht]
\centering
\begin{subfigure}[t]{0.23\textwidth}
\centering
\includegraphics[width=\linewidth]{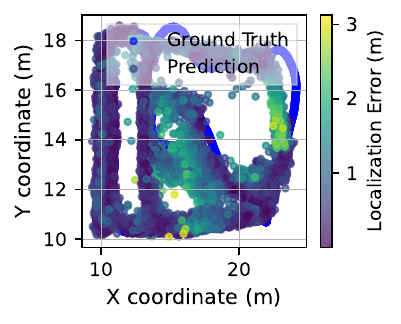}
\caption{\IW.}
\label{fig:scatter}
\end{subfigure}
\hfill
\begin{subfigure}[t]{0.23\textwidth}
\centering
\includegraphics[width=\linewidth]{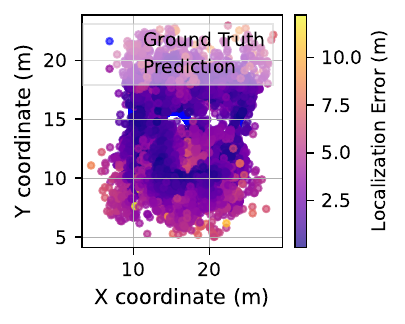}
\caption{Classical TDoA.}
\label{fig:scatter_classical_TDOA}
\end{subfigure}
\vspace{-0.2cm}
\caption{Scatter plot of the predictions versus ground truth positions in the test environment under mixed LOS and NLOS conditions.}
\label{fig:scatter_heatmap}
\vspace{-0.2cm}
\end{figure}

Figure~\ref{fig:scatter_heatmap} provides the localization error for \IW and the classical TDoA approach in the test trajectory that includes both LOS and NLOS conditions to visualize the model estimation quality in the physical space. Regarding \IW (Figure~\ref{fig:scatter}), the majority of points are within a 1\,m error range and cluster very close to their true location, indicating highly \textit{accurate} position estimates. The predictions align closely with the ground truth for most of the route, reflecting errors so small that the points overlap. Only a few points fall within the 2\,m, but these tend to be \textit{isolated}, and there is no systematic drift or offset between predicted and actual tracks. In contrast, the classical TDoA method (Figure~\ref{fig:scatter_classical_TDOA}) shows the opposite pattern, exhibiting \textit{large error} ranges and failing to follow the trajectory of the ground truth. This comparison highlights the superior accuracy and consistency of \IW, particularly in mixed LOS and NLOS conditions, where the classical approach often collapses.
For our security focus, the key takeaway is that \textbf{we did not sacrifice accuracy for security}. In fact, we improved general robustness, which is a prerequisite for security. A system that barely works in normal conditions has no hope against attacks.

\vspace{-0.05cm}
\subsection{Robustness to Anomalous Signals}
\vspace{-0.05cm}
Next, we evaluate the system under conditions simulating adversarial interference. Conducting real RF attacks in a controlled manner is non-trivial, so we use two approaches:  \textbf{(1)} hold-out environment tests, and \textbf{(2)} injection of synthetic perturbations into the data.

\noindent \tikz[baseline=(char.base)]\node[shape=circle,draw,inner sep=0.8pt] (char) {1}; \textbf{Unseen Environment Test.}~Although not a malicious attack per se, the LOS part of the test set can be thought of as a \textit{form of distribution shift}, as the model was trained  only in NLOS regions and then tested under different conditions. Many ML models would degrade under such a shift, similar to how an attack introduces a shift. Our model’s relatively low error in LOS indicates resilience. By comparison, a model variant without anchor positions and TDoA had at least \textit{twice the error} in LOS areas, essentially failing to generalize. This suggests that our model is robust to at least one kind of “new scenario,” which in a security context could translate to not being easily fooled by unexpected channel conditions. It’s not a deliberate attack, but \textit{it builds confidence that the model isn’t 
brittle}.

\noindent \tikz[baseline=(char.base)]\node[shape=circle,draw,inner sep=0.8pt] (char) {2}; \textbf{Spoofing Attack.}~We take a subset of evaluation samples and modify the CIR from a particular anchor to mimic a spoofing scenario. Specifically, we choose the anchor that was given the highest attention weight normally (i.e., the \textit{most influential anchor} for that location) – an intelligent attacker might target that anchor. We then \textit{inject a fake early path} into its CIR: we add a secondary peak to $h_i(\tau)$ at a time corresponding to, for example, 5\,m closer than the true distance, with an amplitude set to 50\% of the main peak. This simulates an attacker \textit{broadcasting a replica} of the device’s signal from a point 5\,m closer to anchor $i$ than the device really is, timing it to arrive slightly before the real direct path. The data from the other anchors remains unchanged.

\begin{figure}[ht]
\centering
\vspace{-0.4cm}
\begin{subfigure}[t]{0.23\textwidth}
\centering
\includegraphics[width=\linewidth]{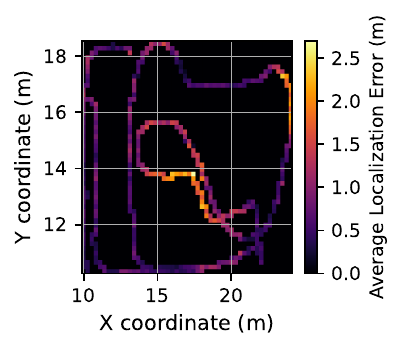}
\caption{\IW with attention.}
\label{fig:heatmap}
\end{subfigure}
\hfill
\begin{subfigure}[t]{0.23\textwidth}
\centering
\includegraphics[width=\linewidth]{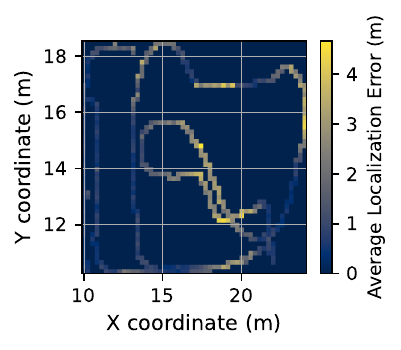}
\caption{\IW without attention.}
\label{fig:heatmap_no-attention}
\end{subfigure}
\vspace{-0.2cm}
\caption{Heatmap of average localization errors for \IW in the test environment under mixed LOS and NLOS conditions, with a spoofing attack targeting the most influential anchor.}
\label{fig:spoofing}
\vspace{-0.35cm}
\end{figure}

We feed this “attacked” data into our trained \IW and measure the resulting localization error. We find that the model’s estimate shifts slightly toward the fake location, but only partially. The mean error increases from 0.58\,m to 0.81\,m when this spoof is present – notably, much less than 5\,m, indicating the model did \textit{not fully rely on the fake signal}. In some instances, the attention weight for anchor $i$ dropped from, e.g., 0.3 to 0.1, meaning the model relied more on other anchors to triangulate. We also ran the same attacked data through a version of the model \textit{without attention mechanism}, where all anchors are weighted equally. In this case, the mean error under attack jumped by about 1.38\,m, and we observed location estimates biased significantly toward the spoofed direction. 
This comparative result is illustrated  in Figure~\ref{fig:spoofing}. In \IW (Figure~\ref{fig:heatmap}), the error remains low, especially in NLOS regions, indicating robust resilience to the spoofing attempt. In contrast, the non-attention variant (Figure~\ref{fig:heatmap_no-attention}) shows significant drift toward the spoofed location, particularly in LOS areas, highlighting that the attention mechanism improves resilience to localized attacks. 

It’s important to note that our model was not explicitly trained on this attack, but its robustness emerges from the model’s design. This is a promising sign for \textbf{zero-shot robustness}, where the system can handle certain attacks it has not seen before.

\noindent \textbf{Other Perturbations.}~We also tested robustness to random noise bursts by adding Gaussian \textit{noise} to the CIR of the most influential anchor, as well as the impact of entirely dropping the anchor to simulate \textit{jamming or signal blockage}. The model effectively handled its absence by naturally relying on the remaining anchors, with the mean error increasing only to 0.93\,m.
Noise bursts had minimal effect unless the SNR was extremely low, in which case that anchor was effectively lost. 

\begin{figure}[ht]
\centering
\begin{subfigure}[t]{0.23\textwidth}
\centering
\includegraphics[width=\linewidth]{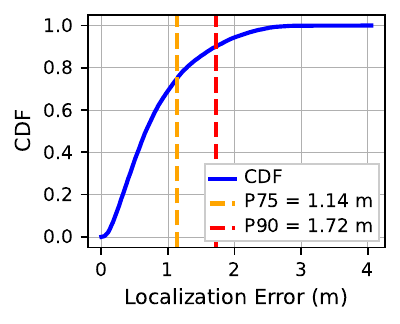}
\caption{$\sigma = 0.2$.}
\label{fig:cdf-sigma0.3}
\end{subfigure}
\hfill
\begin{subfigure}[t]{0.23\textwidth}
\centering
\includegraphics[width=\linewidth]{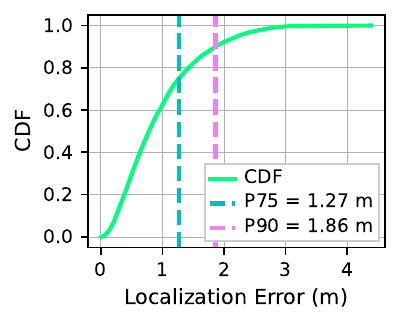}
\caption{$\sigma = 0.5$.}
\label{fig:cdf-sigma0.5}
\end{subfigure}
\vspace{-0.2cm}
\caption{CDFs of localization errors on the mixed LOS and NLOS test trajectory for \IW with Gaussian noise added to the most influential anchor.}
\label{fig:noise}
\vspace{-0.55cm}
\end{figure}

Figure~\ref{fig:noise} presents the cumulative distribution functions (CDFs) of localization errors for \IW under different levels of Gaussian noise added to the most influential anchor. In Figure~\ref{fig:cdf-sigma0.3}, where the standard deviation is set to $\sigma = 0.2$, the model maintains a strong accuracy, with the $p75$ error reaching only 1.14\,m and the $p90$ capped at 1.72\,m. Increasing the noise to $\sigma = 0.5$ (Figure~\ref{fig:cdf-sigma0.5}) slightly degrades performance, with $p75$ reaching 1.27\,m and $p90$ extending to 1.86\,m. Despite the added perturbations, \IW remains resilient, \textit{adjusting attention weights} to reduce the impact of noise.

While these simulations are encouraging, we caution that a determined adversary with knowledge of our model could potentially craft more cunning perturbations. For instance, they might try to fool \textit{multiple anchors simultaneously} in a coordinated way that is still geometrically consistent. Defending against that would likely require explicit adversarial training or anomaly detection beyond what we have. 




\vspace{-0.05cm}
\subsection{Ablation Study}
\vspace{-0.05cm}
We perform an ablation study to quantify the contribution of each component of \IW, which also indirectly validates the security design. We consider three key ablations: \textit{non-attention}, \textit{non-TDoA}, and \textit{non-anchor position and non-TDoA (pure fingerprinting)}. All ablation studies were tested under mixed LOS and NLOS conditions, as shown in Table~\ref{tab: ablation studies}.

\vspace{-0.1cm}
\begin{table}[ht]
\centering
\caption{Performance of \IW against ablation variants in mixed LOS and NLOS conditions.}
\vspace{-0.2cm}
\resizebox{\linewidth}{!}{
\begin{tabular}{llcccc}
\toprule
 & \textbf{Variant} & \textbf{$Mean$} & \textbf{$Median$} & \textbf{$P75$} & \textbf{$P90$} \\
 & \textbf{} & \textbf{$(m)$} & \textbf{$(m)$} & \textbf{$(m)$} & \textbf{$(m)$} \\
\midrule
\multirow{1}{*}{\textbf{\IW}} 
    & All enabled features             & 0.58 & 0.46 & 0.75 & 1.19 \\
\midrule
\multirow{4}{*}{\textbf{Ablations}} 
    & Non-attention                     & 1.37 & 0.89 & 1.81 & 3.07 \\
    & Non-TDoA                          & 1.06 & 0.70 & 1.40 & 2.47 \\
    & Non-anchor position and non-TDoA              & 1.36 & 0.87 & 1.84 & 2.91 \\
\bottomrule
\end{tabular}
}
\vspace{-0.2cm}
\label{tab: ablation studies}
\end{table}

\noindent $\bullet$ \textbf{Non-Attention}.~In this variant, we remove the multi-head attention and simply concatenate all anchor features. The rest of the network is unchanged. We found the performance dropped, with the mean error roughly doubled to 1.37\,m, and $p90$ went to 3.07\,m, compared to 1.19\,m with attention. In particular, many estimates in LOS cases became unstable. Essentially, without attention the model cannot adapt to which anchors are LOS against NLOS at runtime. From a security view, this variant lacks the ability to \textit{discount a compromised anchor}. Indeed, in our one-anchor spoofing simulation, this non-attention model was far more affected. This confirms that attention is a crucial component for both accuracy and robustness.

\noindent $\bullet$ \textbf{Non-TDoA.}~Here, we keep attention but do not feed the relative time inputs. The model only receives anchors’ raw CIR features and attempts to infer the position implicitly. This caused about 50\% higher error (model failed to extrapolate) in LOS and NLOS parts and overall 1.06\,m mean error. It still does better than $k$-NN baseline, indicating the attention mechanism with CNN learns something of geometry implicitly, but it clearly struggled with unseen conditions. This aligns with prior observations that \textit{pure fingerprint models overfit to training conditions}. In terms of attack, a non-TDoA model is easier to fool by creating a fingerprint pattern that it thinks it knows, as it has no physical constraints to check against. Thus, including timing features significantly improves robustness by giving the model a “\textit{reality check}” anchored in known physics.

\noindent $\bullet$ \textbf{Non-Anchor Position and Non-TDoA.}~This extreme variant removes all physical inputs, essentially collapsing \IW into a plain deep fingerprinting model, where multiple CIRs are input and a position is output. Unsurprisingly, this model does not perform well, with a mean error of 1.36\,m and a $p90$ of 2.91\,m. It behaves like a traditional fingerprint system – fine in areas similar to training (NLOS) but fails badly in new scenarios. The model \textit{lacks any mechanism to cope with out-of-distribution signals or attacks}. This is essentially what many naive deep learning localization models would be without security considerations. Our full model’s superiority over this variant, with the \textit{error reduced by half}, quantitatively demonstrates the value of our security-motivated design choices.


\noindent \textbf{Computational Performance:}~Finally, we note that our model is computationally efficient: with approximately 200k parameters, it achieves inference for a single instance (8 anchors) in just a few \textmu{}s on a GPU and about 1\,ms on a CPU. This is negligible compared to a 5G frame interval (e.g., 10\,ms), making the system feasible for real-time tracking. The attention mechanism adds \textit{minimal overhead} and the model could even be deployed on edge devices like smartphone or IoT device to localize itself, which has privacy benefits by \textit{keeping raw CSI data local}. 

\vspace{-0.05cm}
\section{Conclusion and Future Work}
\label{conclusion}
\vspace{-0.05cm}
We presented a security-focused design of a deep learning indoor localization system for 5G networks. By introducing a formal threat model and tailoring the architecture to address spoofing and adversarial perturbations, we demonstrated that it is possible to achieve \textit{accurate and robust} localization in complex environments. Our proposed \IW system combines fingerprinting and geometric approaches, using a CNN with multi-head attention network that inherently filters anomalous signals and enforces physical consistency. It achieved sub-meter accuracy on a public dataset and showed resilience against simulated signal spoofing attacks, significantly outperforming classical methods in mixed conditions LOS and NLOS.


This work takes a step toward \textit{secure and trustworthy indoor positioning}, but it also highlights open challenges. 
Future work will delve into formal adversarial robustness guarantees for RF localization, incorporating anomaly detectors to explicitly flag attacks, and extending the approach to other localization modalities (Wi-Fi, UWB) and multi-floor scenarios. We also plan to explore adversarial training with physically modeled attacks to further harden the system. Another important direction is user privacy, ensuring that the benefits of 5G localization do not come at the expense of exposing sensitive location information to unintended parties, possibly by integrating cryptographic protocols with our ML framework.


\vspace{-0.09cm}
\section*{Acknowledgments}
This research was made possible with the support of the Horizon Europe research and innovation programme of the European Union, under grant agreement number 101092912 (project MLSysOps).
\vspace{-0.09cm}

\bibliographystyle{IEEEtran}
\vspace{-0.09cm}
\bibliography{sample}

\end{document}